\documentclass[12pt]{article}
\usepackage{graphicx}
\begin{document}

\begin{center}
{\Large \bf Harmonic Oscillators as Bridges between Theories: Einstein, Dirac, and
Feynman}

\vspace{3ex}

Y. S. Kim\footnote{electronic address: yskim@physics.umd.edu}\\
Department of Physics, University of Maryland,\\
College Park, Maryland 20742, U.S.A.\\

\vspace{3ex}

Marilyn E. Noz \footnote{electronic address: noz@nucmed.med.nyu.edu}\\
Department of Radiology, New York University,\\ New York, New York 10016, U.S.A.\\

\end{center}

\vspace{3ex}

\begin{abstract}

Other than scattering problems where perturbation theory is applicable,
there are basically two ways to solve problems in physics.  One is
to reduce the problem to harmonic oscillators, and the other is to
formulate the problem in terms of two-by-two matrices.  If two
oscillators are coupled, the problem combines both two-by-two matrices
and harmonic oscillators.  This method then becomes a powerful research
tool to cover many different branches of physics.  Indeed, the concept
and methodology in one branch of physics can be translated into another
through the common mathematical formalism.
Coupled oscillators provide clear illustrative examples for some
of the current issues in physics, including entanglement, decoherence,
and Feynman's rest of the universe.  In addition, it is noted that
the present form of quantum mechanics is largely a physics of harmonic
oscillators.  Special relativity is the physics of the
Lorentz group which can be represented by the group of by two-by-two
matrices commonly called $SL(2,c)$.  Thus the coupled harmonic oscillators
can therefore play the role of combining quantum mechanics with special
relativity.  Both Paul A. M. Dirac and Richard P. Feynman were fond of
harmonic oscillators, while they used different approaches to physical
problems.  Both were also keenly interested in making quantum mechanics
compatible with special relativity.  It is shown that the coupled harmonic
oscillators can bridge these two different approaches to physics.

\end{abstract}

\newpage

\section{Introduction}\label{intro}

Because of its mathematical simplicity, the harmonic oscillator provides
soluble models in many branches of physics.  It often gives a clear
illustration of abstract ideas.  In many cases, the problems are reduced
to the problem of two coupled oscillators.  Soluble models in quantum
field theory, such as the Lee model~\cite{sss61} and the Bogoliubov
transformation in superconductivity~\cite{fewa71}, are based on two
coupled oscillators.    More recently, the coupled oscillators form
the mathematical basis for squeezed states in quantum
optics~\cite{knp91}.

According to our experience, the present form of quantum mechanics
is largely a physics of harmonic oscillators.
Since the group $SL(2,C)$ forms the universal covering group of the
Lorentz group, special relativity is a physics of two-by-two matrices.
Therefore, the coupled harmonic oscillator can provide a concrete model
for relativistic quantum mechanics.

With this point in mind, Dirac and Feynman used harmonic oscillators
to test their physical ideas.  In this paper, we first examine Dirac's
attempts to combine quantum mechanics with relativity in his own style:
to construct mathematically appealing models.  We then examine how
Feynman approached this problem.  He was insisting on his own style.
Observe the experimental world, tell the story of the real world, and
then write down mathematical formulas as needed.

In this paper, we use coupled harmonic oscillators to build a bridge
between the two different attempts made by Dirac and Feynman.  The
coupled oscillator system not only connects the ideas of these two
physicists, but also serves as an illustrative tool for some of the
current ideas in physics, such as entanglement and decoherence.

Feynman's rest of the universe is a case in point.  We shall show in
this paper, using coupled harmonic oscillators, that this concept is
a special case of entanglement.  In their paper
1999 paper~\cite{hkn99ajp} Han {\it et al.} used two coupled harmonic
oscillators to interpret what Feynman said in his book.  There one
oscillator played as the world in which we do physics, and the other
oscillator as the rest of the universe.  We shall see in this paper
that the concept of Feynman's rest of the universe can be expanded to
the concept of entanglement.

Since the same coupled oscillators can be used for both illustrating
entanglement and for the oscillator-based  relativistic quantum
mechanics, we are able to extend the concept of entanglement to the
Lorentz-covariant world.  In so doing, we arrive at the concept of
space-time entanglements.  Indeed, the space-time entanglement is
is one of the essential ingredients in the covariant formulation of
relativistic quantum mechanics.

In Sec.~\ref{quantu}, we start with the classical Hamiltonian for two
coupled oscillators.  It is possible to obtain a explicit solution for
the Schr\"odinger equation in terms of the normal coordinates.  We
then derive a convenient form of this solution from which the concept
of entanglement can be studied thoroughly.
In Sec.~\ref{frest}, we construct the density matrix using the solution
given in Sec.~\ref{quantu}, and explain the effect of the rest of the
universe which we  are not able to observe.
Section~\ref{dirosc} examines Dirac's life-long attempt to combine
quantum mechanics with special relativity.  In Sec.~\ref{adden}, we
study some of the problems which Dirac left us to solve.
In Sec.~\ref{feyosc},
starting from Dirac's work, we construct a covariant model of
relativistic extended particles by combining Dirac's oscillators
with Feynman's phenomenological approach to relativistic quark model.
It is shown that Feynman's parton model can be interpreted as a limiting
case of one covariant model for a covariant bound-state model.

\section{Coupled Oscillators and Entangled Oscillators}\label{quantu}

Two coupled harmonic oscillators serve many different purposes in
physics.  It is well known that this oscillator problem can be
formulated into a problem of a quadratic equation in two variables.
The diagonalization of the quadratic form includes a rotation of the
coordinate system.  However, the diagonalization process requires
additional transformations involving the scales of the coordinate
variables~\cite{hkn99ajp,arav89}.  Indeed, it was found that the
mathematics of this procedure can be as complicated as the group
theory of Lorentz transformations in a six dimensional space with
three spatial and three time coordinates~\cite{hkn95jm}.

However, in this paper, we start with a simple problem of two oscillators
with equal mass.  This contains enough physics for our present purpose.
Then the Hamiltonian takes the form
\begin{equation}\label{eq.1}
H = {1\over 2}\left\{{1\over m} p^{2}_{1} + {1\over m}p^{2}_{2}
+ A x^{2}_{1} + A x^{2}_{2} + 2C x_{1} x_{2} \right\}.
\end{equation}

If we choose coordinate variables
\begin{eqnarray} \label{eq.3}
&{}& y_{1} = {1\over\sqrt{2}}\left(x_{1} + x_{2}\right) , \nonumber\\[2ex]
&{}& y_{2} = {1\over\sqrt{2}}\left(x_{1} - x_{2}\right) ,
\end{eqnarray}
the Hamiltonian can be written as
\begin{equation}\label{eq.6}
H = {1\over 2m} \left\{p^{2}_{1} + p^{2}_{2} \right\} +
{K\over 2}\left\{e^{-2\eta} y^{2}_{1} + e^{2\eta} y^{2}_{2} \right\} ,
\end{equation}
where
\begin{eqnarray}\label{eq.5}
&{}&   K = \sqrt{A^{2} - C^{2}} ,  \nonumber \\[.5ex]
&{}& \exp(2\eta) =\sqrt{\frac{A - C}{A + C} } ,
\end{eqnarray}
The classical eigenfrequencies are $\omega_{pm} = \omega e^{\pm}$ with
\begin{equation}\label{omega}
\omega = \sqrt{\frac{K}{m}} .
\end{equation}

If $y_{1}$ and $y_{2}$ are measured in units of $(mK)^{1/4} $,
the ground-state wave function of this oscillator system is
\begin{equation}\label{eq.13}
\psi_{\eta}(x_{1},x_{2}) = {1 \over \sqrt{\pi}}
\exp{\left\{-{1\over 2}(e^{-\eta} y^{2}_{1} + e^{\eta} y^{2}_{2})
\right\} } ,
\end{equation}
The wave function is separable in the $y_{1}$ and $y_{2}$ variables.
However, for the variables $x_{1}$ and $x_{2}$, the story is quite
different, and can be extended to the issue of entanglement.

There are three ways to excite this ground-state oscillator system.
One way is to multiply Hermite polynomials for the usual quantum
excitations.  The second way is to construct coherent states for
each of the $y$ variables.  Yet, another way is to construct
thermal excitations.  This requires density matrices and Wigner
functions~\cite{hkn99ajp}.

The key question is how the quantum mechanics in the world of the
$x_{1}$ variable is affected by the $x_{2}$ variable.  If the
$x_{2}$ space is not observed, it corresponds to Feynman's rest
of the universe.  If we use two separate measurement processes for
these two variables, these two oscillators are  entangled.

Let us write the wave function of Eq.(\ref{eq.13}) in terms of
$x_{1}$ and $x_{2}$, then
\begin{equation}\label{eq.14}
\psi_{\eta}(x_{1},x_{2}) = {1 \over \sqrt{\pi}}
\exp\left\{-{1\over 4}\left[e^{-\eta}(x_{1} + x_{2})^{2} +
e^{\eta}(x_{1} - x_{2})^{2} \right] \right\} .
\end{equation}
When the system is decoupled with $\eta = 0$, this wave function becomes
\begin{equation}\label{eq.15}
\psi_{0}(x_{1},x_{2}) = \frac{1}{\sqrt{\pi}}
\exp{\left\{-{1\over 2}(x^{2}_{1} + x^{2}_{2}) \right\}} .
\end{equation}
The system becomes separable and becomes disentangled.

As was discussed in the literature for several different
purposes~\cite{knp91,kno79ajp,knp86}, this wave function can be
expanded as
\begin{equation}\label{expan}
\psi_{\eta }(x_{1},x_{2}) = {1 \over \cosh\eta}\sum^{}_{k}
\left(\tanh{\eta \over 2}\right)^{k} \phi_{k}(x_{1}) \phi_{k}(x_{2}) ,
\end{equation}
where $\phi_{k}(x)$ is the harmonic oscillator wave function for the
$k-th$ excited state.
This expansion serves as the mathematical basis for squeezed states
of light in quantum optics~\cite{knp91}, among other applications.

In addition, this expression clearly demonstrates that the coupled
oscillators are entangled oscillators.  Let us look at the expression
of Eq.(\ref{expan}).  If the variable $x_{1}$ and $x_{2}$ are measured
separately.

In Sec~\ref{dirosc}, we shall see that the mathematics of the coupled
oscillators can serve as the basis for the covariant harmonic
oscillator formalism where the $x_{1}$ and $x_{2}$ variables
are replaced by the longitudinal and time-like variables,
respectively.  This mathematical identity will leads to the
concept of space-time entanglement in special relativity.

\section{Feynman's Rest of the Universe}\label{frest}

In his book on statistical
mechanics~\cite{fey72}, Feynman makes the following statement about the
density matrix. {\it When we solve a quantum-mechanical problem, what we
really do is divide the universe into two parts - the system in which we
are interested and the rest of the universe.  We then usually act as if
the system in which we are interested comprised the entire universe.
To motivate the use of density matrices, let us see what happens when we
include the part of the universe outside the system}.

We can use the coupled harmonic oscillators to illustrate what Feynman
says in his book.  Here we can use $x_{1}$ and $x_{2}$ for the variable
we observe and the variable in the rest of the universe.  By using the
rest of the universe, Feynman does not rule out the possibility of other
creatures measuring the $x_{2}$ variable in their part of the universe.

Using the wave function $\psi_{\eta}(x_{1},x_{2})$ of Eq.(\ref{eq.14}),
we can construct the pure-state density matrix
\begin{equation}
\rho(x_{1},x_{2};x_{1}',x_{2}')
= \psi_{\eta}(x_{1},x_{2})\psi_{\eta}(x_{1}',x_{2}') ,
\end{equation}
which satisfies the condition $\rho^{2} = \rho $:
\begin{equation}
\rho(x_{1},x_{2};x_{1}',x_{2}') =
\int \rho(x_{1},x_{2};x_{1}'',x_{2}'')
\rho(x_{1}'',x_{2}'';x_{1}',x_{2}') dx_{1}'' dx_{2}'' .
\end{equation}
If we are not able to make observations on $x_{2}$, we should
take the trace of the $\rho$ matrix with respect to the $x_{2}$
variable.  Then the resulting density matrix is
\begin{equation}\label{integ}
\rho(x_{1}, x_{1}') = \int \rho (x_{1},x_{2};x'_{1},x_{2}) dx_{2} .
\end{equation}

The above density matrix can also be calculated from the expansion of
the wave function given in Eq.(\ref{expan}).  If we perform the integral
of Eq.(\ref{integ}), the result is
\begin{equation}\label{dmat}
\rho(x,x') = \left({1 \over \cosh(\eta/2)}\right)^{2}
\sum^{}_{k} \left(\tanh{\eta \over 2}\right)^{2k}
\phi_{k}(x)\phi^{*}_{k}(x') ,
\end{equation}
where we now use $x$ and $x'$ for $x_{1}$ and $x_{1}'$ for simplicity.
The trace of this density matrix is $1$.  It is also straightforward to
compute the integral for $Tr(\rho^{2})$.  The calculation leads to
\begin{equation}
Tr\left(\rho^{2} \right)
= \left({1 \over \cosh(\eta/2)}\right)^{4}
\sum^{}_{k} \left(\tanh{\eta \over 2}\right)^{4k} .
\end{equation}
The sum of this series is $(1/\cosh\eta)$ which is less than one.

This is of course due to the fact that we are averaging over the $x_{2}$
variable which we do not measure.  The standard way to measure this
ignorance is to calculate the entropy defined as \cite{wiya63}
\begin{equation}
S = - Tr\left(\rho \ln(\rho) \right) ,
\end{equation}
where $S$ is measured in units of Boltzmann's constant.  If we use the
density matrix given in Eq.(\ref{dmat}), the entropy becomes
\begin{equation}
S = 2 \left\{\cosh^{2}\left({\eta \over 2}\right)
\ln\left(\cosh{\eta \over 2}\right) -
\sinh^{2}\left({\eta \over 2}\right)
\ln\left(\sinh{\eta \over 2} \right)\right\} .
\end{equation}
This expression can be translated into a more familiar form if
we use the notation
\begin{equation}
\tanh{\eta \over 2} = \exp\left(-{\hbar\omega \over kT}\right) ,
\end{equation}
where $\omega$ is given in Eq.(\ref{omega}). The ratio
$\hbar\omega/kT$ is a dimensionless variable.  In terms of
this variable, the entropy takes the form
\begin{equation}
S = \left({\hbar\omega \over kT}\right)
\frac{1}{\exp(\hbar\omega/kT) - 1}
- \ln\left[1 - \exp(-\hbar\omega/kT)\right] .
\end{equation}
This familiar expression is for the entropy of an oscillator state in
thermal equilibrium.  Thus, for this oscillator system, we can relate
our ignorance to the temperature.  It is interesting to note that the
coupling strength measured by $\eta$ can be related to the temperature
variable.

\section{Dirac's Harmonic Oscillators}\label{dirosc}

Paul A. M. Dirac is known to us through the Dirac equation for spin-1/2
particles.  But his main interest was in the foundational problems.
First, Dirac was never satisfied with the probabilistic formulation of
quantum mechanics.  This is still one of the hotly debated subjects in
physics.  Second, if we tentatively accept the present form
of quantum mechanics, Dirac was insisting that it has to be consistent
with special relativity.  He wrote several important papers on this
subject.  Let us look at some of his papers on this subject.

\begin{figure}[thb]
\centerline{\includegraphics[scale=0.8]{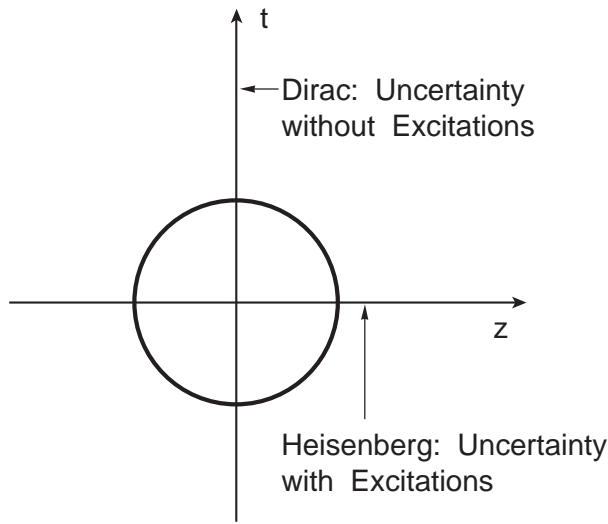}}
\vspace{5mm}
\caption{Space-time picture of quantum mechanics.  There
are quantum excitations along the space-like longitudinal direction, but
there are no excitations along the time-like direction.  The time-energy
relation is a c-number uncertainty relation.}\label{quantum}
\end{figure}

During World War II, Dirac was looking into the possibility of constructing
representations of the Lorentz group using harmonic oscillator wave
functions~\cite{dir45}.  The Lorentz group is the language of special
relativity, and the present form of quantum mechanics starts with harmonic
oscillators.  Presumably, therefore, he was interested in making quantum
mechanics Lorentz-covariant by constructing representations of the Lorentz
group using harmonic oscillators.

In his 1945 paper~\cite{dir45}, Dirac considers the Gaussian form
\begin{equation}
\exp\left\{- {1 \over 2}\left(x^2 + y^2 + z^2 + t^2\right)\right\} .
\end{equation}
We note that this Gaussian form is  in the $(x,~y,~z,~t)$
coordinate variables.  Thus, if we consider Lorentz boost along the
$z$ direction, we can drop the $x$ and $y$ variables, and write the
above equation as
\begin{equation}\label{ground}
\exp\left\{- {1 \over 2}\left(z^2 + t^2\right)\right\} .
\end{equation}
This is a strange expression for those who believe in Lorentz invariance.
The expression
\begin{equation}
\exp\left\{- {1 \over 2}\left(z^2 - t^2\right)\right\} .
\end{equation}
is invariant, but Dirac's Gaussian form of Eq.(\ref{ground}) is not.

On the other hand, this expression is consistent with his earlier papers
on the time-energy uncertainty relation~\cite{dir27}.  In those papers,
Dirac observes that there is a time-energy uncertainty relation, while
there are no excitations along the time axis.  He called this the
``c-number time-energy uncertainty'' relation.  When one of us
(YSK) was talking with Dirac in 1978, he clearly mentioned
this word again.  He said further that this is one of the stumbling
block in combining quantum mechanics with relativity.  This
situation is illustrated in Fig.~\ref{quantum}.

\begin{figure}[thb]
\centerline{\includegraphics[scale=0.8]{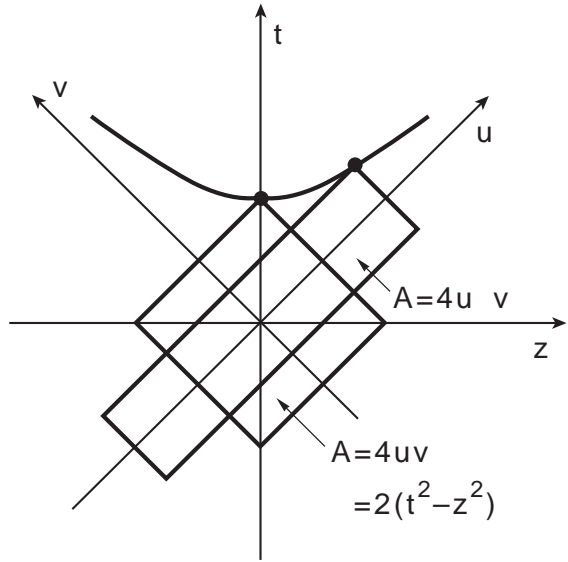}}
\vspace{5mm}
\caption{Lorentz boost in the light-cone coordinate
system.}\label{licone}
\end{figure}

Let us look at Fig.~\ref{quantum} carefully.  This figure is a pictorial
representation of Dirac's Eq.(\ref{ground}),  with localization in both
space and time coordinates.  Then Dirac's fundamental question would be
how to make this figure covariant?  This is where Dirac stops.  However,
this is not the end of the Dirac story.
\begin{figure}[thb]
\centerline{\includegraphics[scale=0.4]{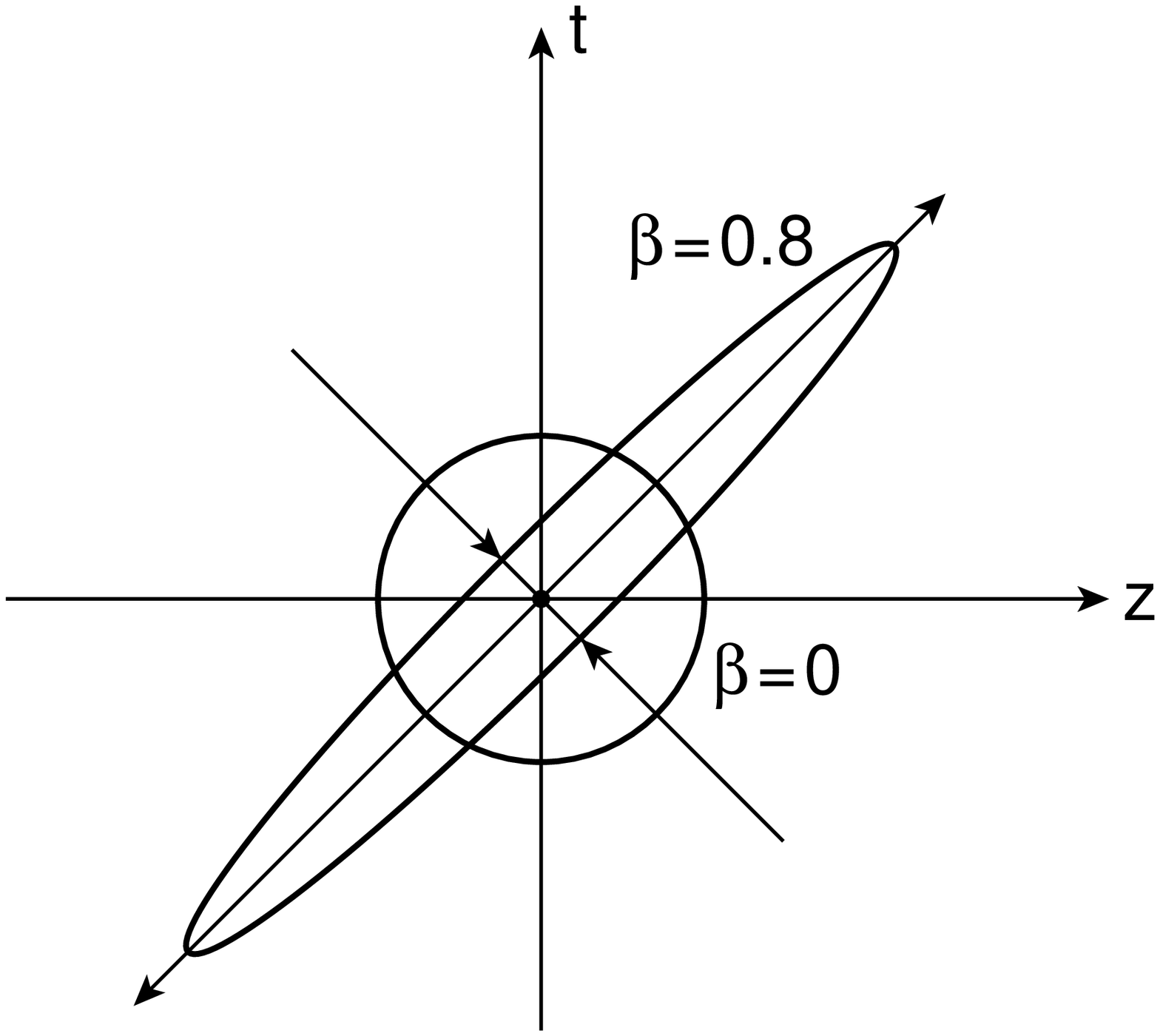}}
\caption{Effect of the Lorentz boost on the space-time
wave function.  The circular space-time distribution in the rest frame
becomes Lorentz-squeezed to become an elliptic
distribution.}\label{ellipse}
\end{figure}

Dirac's interest in harmonic oscillators did not stop with his 1945
paper on the representations of the Lorentz group.  In his
1963~\cite{dir63} paper, he constructed a representation
of the $O(3,2)$ deSitter group using two coupled harmonic oscillators.
This paper contains not only the mathematics of combining special
relativity with the quantum mechanics of quarks inside hadrons, but
also forms the foundations of two-mode squeezed states which are so
essential modern quantum optics~\cite{knp91}.   Dirac did not know
these when he was writing this 1963 paper.

Furthermore, the $O(3,2)$ deSitter group contains the Lorentz group
$O(3,1)$ as a subgroup.  Thus, Dirac's oscillator representation of
the deSitter group essentially contains all the mathematical
ingredient of what we are doing in this paper.

\section{Addendum to Dirac'c Oscillators}\label{adden}

In 1949, the Reviews of Modern Physics published a special issue to
celebrate Einstein's 70th birthday.  This issue contains Dirac paper
entitled ``Forms of Relativistic Dynamics''~\cite{dir49}.
In this paper, he introduced his light-cone coordinate system,
in which a Lorentz boost becomes a squeeze transformation.

When the system is boosted along the $z$ direction, the transformation
takes the form
\begin{equation}\label{boostm}
\pmatrix{z' \cr t'} = \pmatrix{\cosh(\eta/2) & \sinh(\eta/2) \cr
\sinh(\eta/2) & \cosh(\eta/2) } \pmatrix{z \cr t} .
\end{equation}

This is not a rotation, and people still feel strange about this
form of transformation.  In 1949~\cite{dir49}, Dirac introduced his
light-cone
variables defined as~\cite{dir49}
\begin{equation}\label{lcvari}
u = (z + t)/\sqrt{2} , \qquad v = (z - t)/\sqrt{2} ,
\end{equation}
the boost transformation of Eq.(\ref{boostm}) takes the form
\begin{equation}\label{lorensq}
u' = e^{\eta/2 } u , \qquad v' = e^{-\eta/2 } v .
\end{equation}
The $u$ variable becomes expanded while the $v$ variable becomes
contracted, as is illustrated in Fig.~\ref{licone}.  Their product
\begin{equation}
uv = {1 \over 2}(z + t)(z - t) = {1 \over 2}\left(z^2 - t^2\right)
\end{equation}
remains invariant.  In Dirac's picture, the Lorentz boost is a
squeeze transformation.

If we combine Fig.~\ref{quantum} and Fig.~\ref{licone}, then we end up
with Fig.~\ref{ellipse}.
In mathematical formulae, this transformation changes the Gaussian form
of Eq.(\ref{ground}) into
\begin{equation}\label{eta}
\psi_{\eta }(z,t) = \left({1 \over \pi }\right)^{1/2}
\exp\left\{-{1\over 2}\left(e^{-\eta }u^{2} +
e^{\eta}v^{2}\right)\right\} .
\end{equation}
Let us go back to Sec.~\ref{quantu} on the coupled oscillators.  The
above expression is the same as Eq.(\ref{eq.13}).  The $x_{1}$ variable
now became the longitudinal variable $z$, and the $x_{2}$ variable
became the time like variable $t$.

We can use the coupled harmonic oscillators as the starting point of
relativistic quantum mechanics.  This allows us to translate the quantum
mechanics of two coupled oscillators defined over the space of
$x_{1}$ and $x_{2}$ into the quantum mechanics defined over the
space time region of $z$ and $t$.

This form becomes (\ref{ground}) when $\eta$ becomes zero.  The
transition from Eq.(\ref{ground}) to Eq.(\ref{eta}) is a squeeze
transformation.
It is now possible to combine what Dirac observed into a covariant formulation
of harmonic oscillator system. First, we can combine his c-number
time-energy uncertainty relation described in Fig.~\ref{quantum}
and his light-cone coordinate system of Fig.~\ref{licone} into
a picture of covariant space-time localization given in
Fig.~\ref{ellipse}.

In addition, there are two more homework problems which Dirac left
us to solve. First, in defining the $t$ variable for the Gaussian form of
Eq.(\ref{ground}),  Dirac did not specify the physics of this variable.
If it is going to be the calendar time, this form vanishes in the remote
past and remote future.  We are not dealing with this kind of object in
physics.  What is then the physics of this time-like $t$ variable?

\begin{figure}[thb]
\centerline{\includegraphics[scale=0.7]{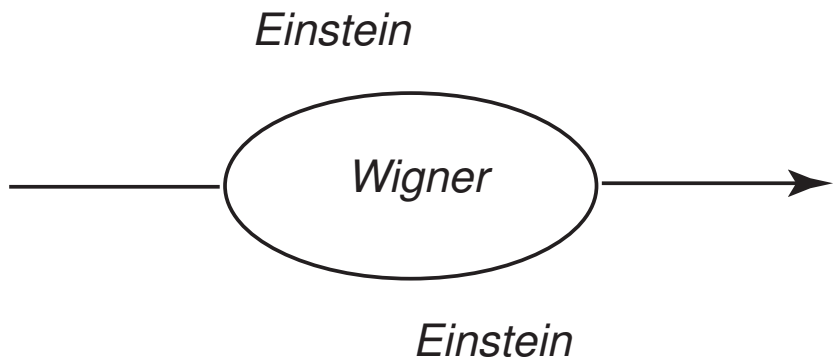}}
\vspace{5mm}
\caption{Wigner in Einstein's world.  Einstein formulates special
relativity whose energy-momentum relation is valid for point particles
as well as particles with internal space-time structure.  It was Wigner
who formulated the framework for internal space-time symmetries by introducing his
little groups whose transformations leave the four-momentum of a given particle
invariant.}\label{dff22}
\end{figure}
The Schr\"odinger quantum mechanics of the hydrogen atom deals with
localized probability distribution.  Indeed, the localization condition
leads to the discrete energy spectrum.  Here, the uncertainty relation
is stated in terms of the spatial separation between the proton and
the electron.  If we believe in Lorentz covariance, there must also
be a time-separation between the two constituent particles, and an
uncertainty relation applicable to this separation variable.  Dirac
did not say in his papers of 1927 and 1945, but Dirac's ``t'' variable
is applicable to this time-separation variable.  This time-separation
variable will be discussed in detail in Sec.~\ref{feyosc} for the
case of relativistic extended particles.

Second, as for the time-energy uncertainty relation.  Dira'c concern
was how the c-cnumber time-energy uncertainty relation without excitations
can be combined with uncertainties in the position space with excitations.
Dira's 1927 paper was written before Wigner's 1939 paper on the internal
space-time symmetries of relativistic particles.

Both of these questions can be answered in terms of the space-time
symmetry of bound states in the Lorentz-covariant regime.  In his
1939 paper, Wigner worked out internal space-time symmetries of
relativistic particles.  He approached the problem by constructing
the maximal subgroup of the Lorentz group whose transformations leave
the given four-momentum invariant.  As a consequence, the internal
symmetry of a massive particle is like the three-dimensional rotation
group.

If we extend this concept to relativistic bound states, the space-time
asymmetry which Dirac observed in 1927 is quite consistent with Einstein's
Lorentz covariance.  The time variable can be treated separately.
Furthermore, it is possible to construct a representations of Wigner's
little group for massive particles~\cite{knp86}.
As for the time-separation, it is also a variable governing
internal space-time symmetry which can be linearly mixed when the
system is Lorentz-boosted.

\section{Feynman's Oscillators }\label{feyosc}

\begin{figure}[thb]
\centerline{\includegraphics[scale=0.7]{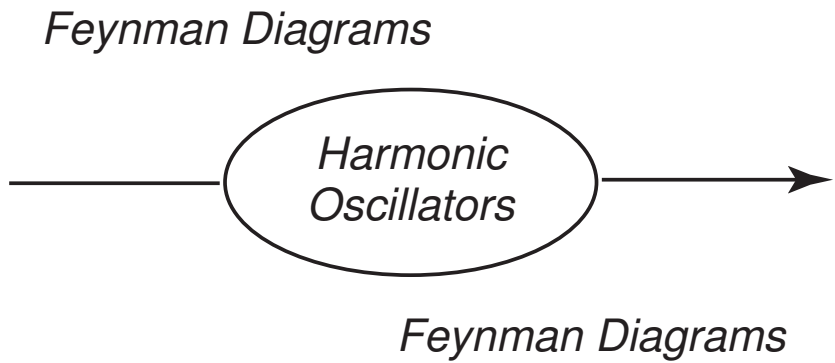}}
\vspace{5mm}
\caption{Feynman's roadmap for combining quantum mechanics with special
relativity.  Feynman diagrams work for running waves, and they provide
a satisfactory resolution for scattering states in Einstein's world.
For standing waves trapped inside an extended hadron, Feynman suggested
harmonic oscillators as the first step.}\label{dff33}
\end{figure}

Quantum field theory has been quite successful in terms of Feynman
diagrams based on the S-matrix formalism, but is useful only for physical
processes where a set of free particles becomes another set of free
particles after interaction.  Quantum field theory does not address the
question of localized probability distributions and their covariance
under Lorentz transformations.  In order to address this question,
Feynman {\it et al.} suggested harmonic oscillators to tackle the
problem~\cite{fkr71}.  Their idea is indicated in Fig.~\ref{dff33}.

Before 1964~\cite{gell64}, the hydrogen atom was used for
illustrating bound states.  These days, we use hadrons which are
bound states of quarks.  Let us use the simplest hadron consisting of
two quarks bound together with an attractive force, and consider their
space-time positions $x_{a}$ and $x_{b}$, and use the variables
\begin{equation}
X = (x_{a} + x_{b})/2 , \qquad x = (x_{a} - x_{b})/2\sqrt{2} .
\end{equation}
The four-vector $X$ specifies where the hadron is located in space and
time, while the variable $x$ measures the space-time separation
between the quarks.  According to Einstein, this space-time separation
contains a time-like component which actively participates as in
Eq.(\ref{boostm}), if the hadron is boosted along the $z$ direction.
This boost can be conveniently described by the light-cone variables
defined in Eq(\ref{lcvari}).
Does this time-separation variable exist when the hadron is at rest?
Yes, according to Einstein.  In the present form of quantum mechanics,
we pretend not to know anything about this variable.  Indeed, this
variable belongs to Feynman's rest of the universe.

What do Feynman {\it et al.} say about this oscillator wave function?
In their classic 1971 paper~\cite{fkr71}, Feynman {\it et al.} start
with the following Lorentz-invariant differential equation.
\begin{equation}\label{osceq}
{1\over 2} \left\{x^{2}_{\mu} -
{\partial^{2} \over \partial x_{\mu }^{2}}
\right\} \psi(x) = \lambda \psi(x) .
\end{equation}
This partial differential equation has many different solutions
depending on the choice of separable variables and boundary conditions.
Feynman {\it et al.} insist on Lorentz-invariant solutions which are
not normalizable.  On the other hand, if we insist on normalization,
the ground-state wave function takes the form of Eq.(\ref{ground}).
It is then possible to construct a representation of the
Poincar\'e group from the solutions of the above differential
equation~\cite{knp86}.  If the system is boosted, the wave function
becomes given in Eq.(\ref{eta}).

\begin{figure}
\centerline{\includegraphics[scale=0.5]{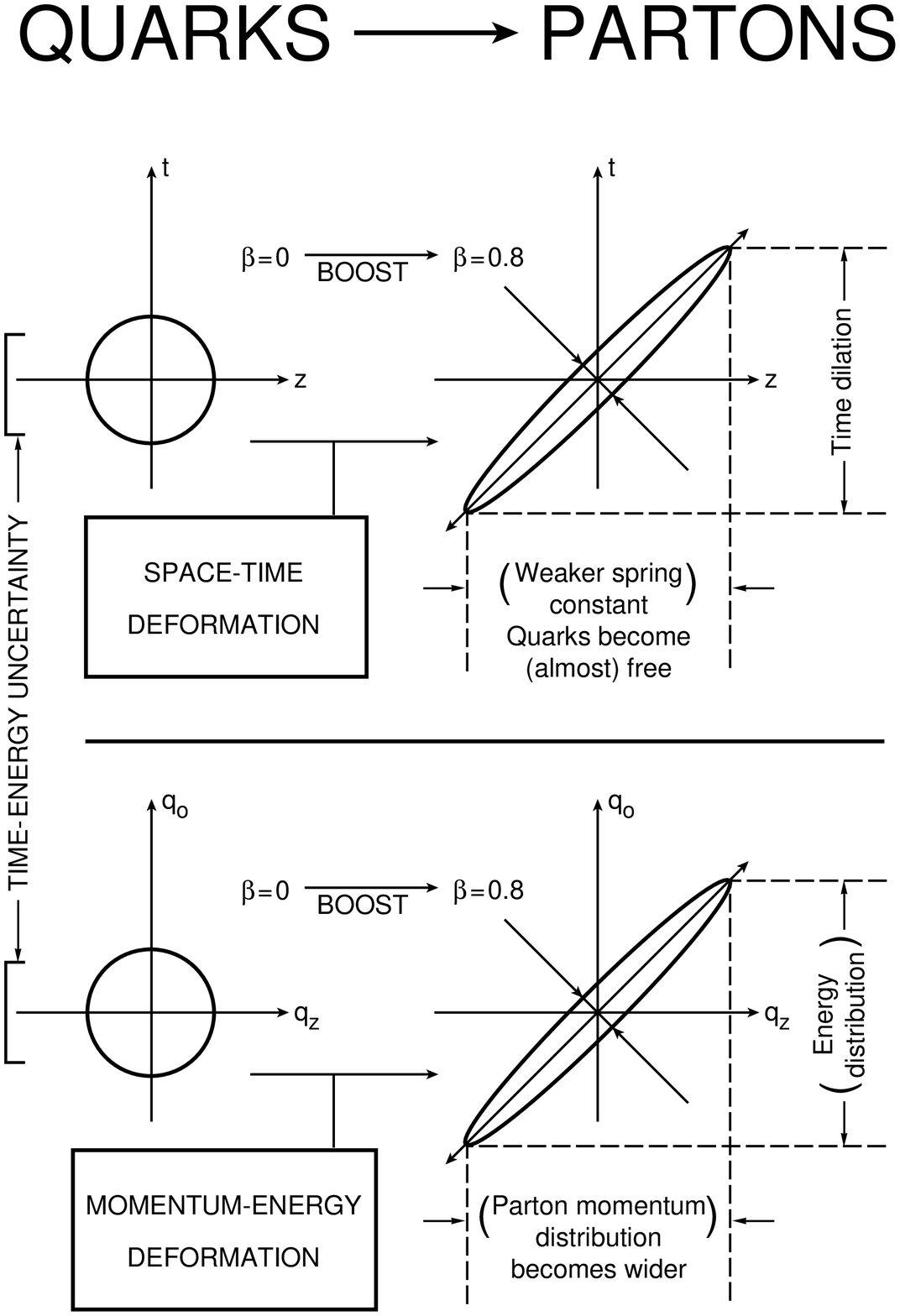}}
\vspace{5mm}
\caption{Lorentz-squeezed space-time and momentum-energy wave
functions.  As the hadron's speed approaches that of light, both
wave functions become concentrated along their respective positive
light-cone axes.  These light-cone concentrations lead to Feynman's
parton picture.}\label{parton}
\end{figure}

This wave function becomes Eq.(\ref{ground}) if $\eta$ becomes zero.
The transition from Eq.(\ref{ground}) to Eq.(\ref{eta}) is a
squeeze transformation.  The wave function of Eq.(\ref{ground}) is
distributed within a circular region in the $u v$ plane, and thus
in the $z t$ plane.  On the other hand, the wave function of
Eq.(\ref{eta}) is distributed in an elliptic region with the light-cone
axes as the major and minor axes respectively.  If $\eta$ becomes very
large, the wave function becomes concentrated along one of the
light-cone axes.  Indeed, the form given in Eq.(\ref{eta}) is a
Lorentz-squeezed wave  function.  This squeeze mechanism is
illustrated in Fig.~\ref{ellipse}.

There are many different solutions of the Lorentz invariant differential
equation of Eq.(\ref{osceq}).  The solution given in Eq.(\ref{eta})
is not Lorentz invariant but is covariant.  It is normalizable in the
$t$ variable, as well as in the space-separation variable $z$.  It is
indeed possible to construct Wigner's $O(3)$-like little group for massive
particles~\cite{wig39}, and thus the representation of the Poincar\'e
group~\cite{knp86}.  Our next question is whether this formalism has
anything to do with the real world.

\begin{figure}
\centerline{\includegraphics[scale=0.6]{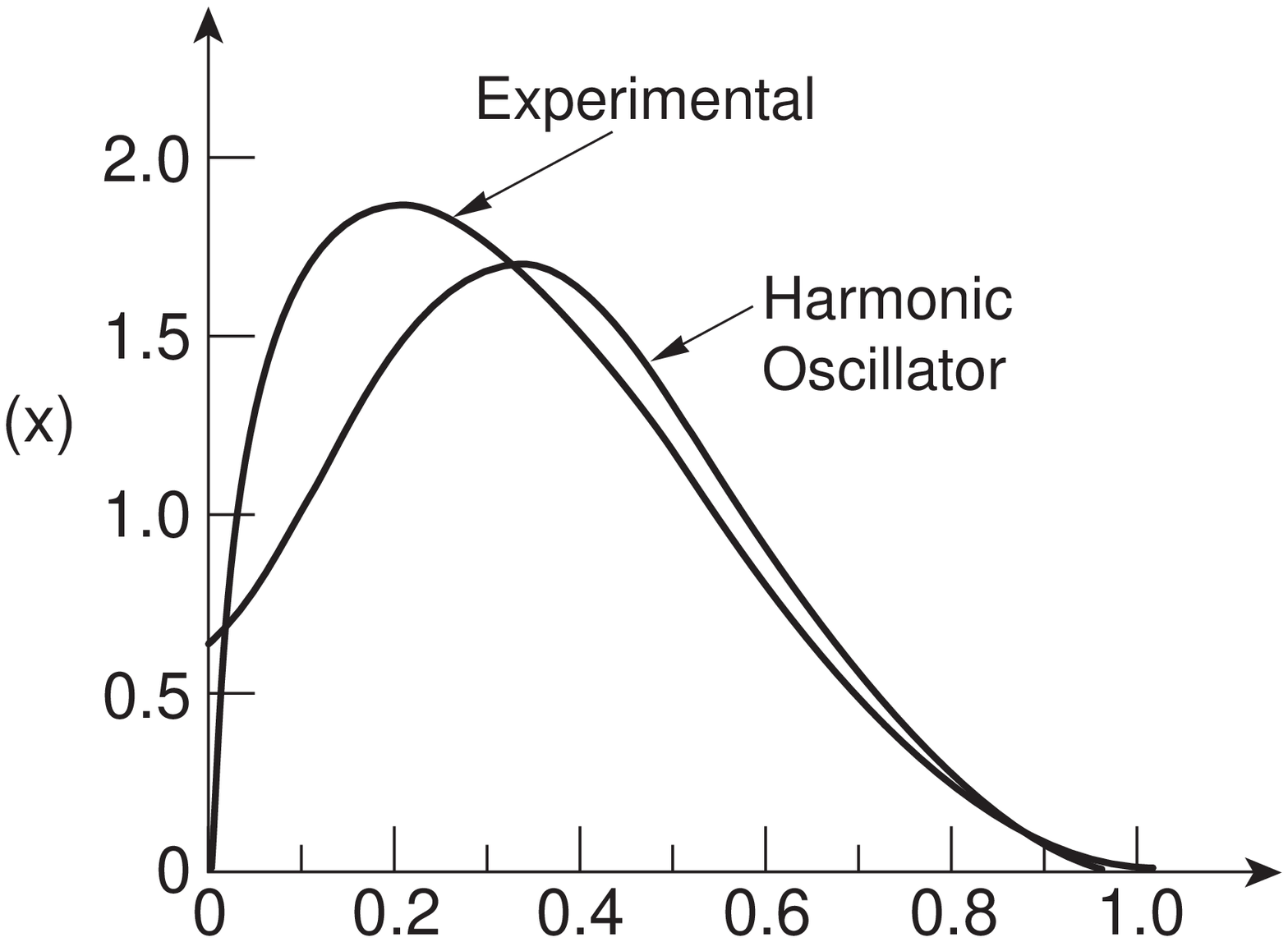}}
\vspace{5mm}
\caption{Parton distribution function.
Theory and experiment.}\label{hussar}
\end{figure}

In 1969, Feynman observed that a fast-moving hadron can be regarded
as a collection of many ``partons'' whose properties appear to be
quite different from those of the quarks~\cite{fey69}.  For example,
the number of quarks inside a static proton is three, while the number
of partons in a rapidly moving proton appears to be infinite.  The
question then is how the proton looking like a bound state of quarks
to one observer can appear different to an observer in a different
Lorentz frame?  Feynman made the following systematic observations.

\begin{itemize}

\item[a.]  The picture is valid only for hadrons moving with
  velocity close to that of light.

\item[b.]  The interaction time between the quarks becomes dilated,
   and partons behave as free independent particles.

\item[c.]  The momentum distribution of partons becomes widespread as
   the hadron moves fast.

\item[d.]  The number of partons seems to be infinite or much larger
    than that of quarks.

\end{itemize}

\noindent Because the hadron is believed to be a bound state of two
or three quarks, each of the above phenomena appears as a paradox,
particularly b) and c) together.

In order to resolve this paradox, let us write down the
momentum-energy wave function corresponding to Eq.(\ref{eta}).
If we let the quarks have the four-momenta $p_{a}$ and $p_{b}$, it is
possible to construct two independent four-momentum
variables~\cite{fkr71}
\begin{equation}
P = p_{a} + p_{b} , \qquad q = \sqrt{2}(p_{a} - p_{b}) ,
\end{equation}
where $P$ is the total four-momentum.  It is thus the hadronic
four-momentum.

The variable $q$ measures the four-momentum separation between
the quarks.  Their light-cone variables are
\begin{equation}\label{conju}
q_{u} = (q_{0} - q_{z})/\sqrt{2} ,  \qquad
q_{v} = (q_{0} + q_{z})/\sqrt{2} .
\end{equation}
The resulting momentum-energy wave function is
\begin{equation}\label{phi}
\phi_{\eta }(q_{z},q_{0}) = \left({1 \over \pi }\right)^{1/2}
\exp\left\{-{1\over 2}\left(e^{\eta}q_{u}^{2} +
e^{-\eta}q_{v}^{2}\right)\right\} .
\end{equation}
Because we are using here the harmonic oscillator, the mathematical
form of the above momentum-energy wave function is identical to that
of the space-time wave function.  The Lorentz squeeze properties of
these wave functions are also the same.  This aspect of the squeeze
has been exhaustively discussed in the
literature~\cite{knp86,kn77par,kim89}.

When the hadron is at rest with $\eta = 0$, both wave functions
behave like those for the static bound state of quarks.  As $\eta$
increases, the wave functions become continuously squeezed until
they become concentrated along their respective positive
light-cone axes.  Let us look at the z-axis projection of the
space-time wave function.  Indeed, the width of the quark distribution
increases as the hadronic speed approaches that of the speed of
light.  The position of each quark appears widespread to the observer
in the laboratory frame, and the quarks appear like free particles.

The momentum-energy wave function is just like the space-time wave
function, as is shown in Fig.~\ref{parton}.  The longitudinal momentum
distribution becomes wide-spread as the hadronic speed approaches the
velocity of light.  This is in contradiction with our expectation from
non-relativistic quantum mechanics that the width of the momentum
distribution is inversely proportional to that of the position wave
function.  Our expectation is that if the quarks are free, they must
have their sharply defined momenta, not a wide-spread distribution.

However, according to our Lorentz-squeezed space-time and
momentum-energy wave functions, the space-time width and the
momentum-energy width increase in the same direction as the hadron
is boosted.  This is of course an effect of Lorentz covariance.
This indeed is the key to the resolution of the quark-parton
paradox~\cite{knp86,kn77par}.

After these qualitative arguments, we are interested in whether
Lorentz-boosted bound-state wave functions in the hadronic rest
frame could lead to parton distribution functions.  If we start with
the ground-state Gaussian wave function for the three-quark wave
function for the proton, the parton distribution function appears
as Gaussian as is indicated in Fig.~\ref{hussar}.  This Gaussian  form
is compared with experimental distribution also in Fig.~\ref{hussar}.

For large $x$ region, the agreement is excellent, but the agreement is
not satisfactory for small values of $x$.  In this region, there is
a complication called the ``sea quarks.''  However, good sea-quark physics
starts from good valence-quark physics.  Figure~\ref{hussar} indicates
that the boosted ground-state wave function provides a good valence-quark
physics.

Feynman's parton picture is one of the most controversial models
proposed in the 20th century.  The original model is valid only in
Lorentz frames where the initial proton moves with infinite momentum.
It is gratifying to note that this model can be produced as a limiting
case of one covariant model which produces the quark model in the
frame where the proton is at rest.

\section*{Concluding Remarks}\label{concl}

The major strength of the coupled oscillator system is that its
classical mechanics is known to every physicist.  Not too well known
is the fact that this simple device can serve as an analog computer
for many of the current problems in physics.

This oscillator system was very useful in illustrating Feynman's
rest of the universe~\cite{hkn99ajp}.  In this report, we have shown
first that the coupled oscillator system can server as an illustrative
example of the concept of entanglement, and that Feynman's rest of
the universe is a special case of entanglement.  Conversely, the
the rest of the universe can be extended to the concept of
entanglement.

It was also noted that the coupled-oscillator system provides the
mathematical basis for the covariant harmonic oscillators.  It can
also translate the problems of entanglement to the space and time
variables.

It is well known that harmonic oscillators provide bridges between
theories.  In this paper, we have seen that the coupled harmonic
oscillators can serve as a bridge between Dirac and Feynman, and
a bridge between coupled oscillators and harmonic oscillators in
the Lorentz-covariant world.

\section*{Acknowledgments}
We would like to thank G. S. Agarwal, H. Hammer, and A. Vourdas for
helpful discussion on the precise definition of the word ``entanglement''
applicable to coupled systems.

\end{document}